\documentclass[prd,twocolumn,amsmath,amssymb,floatfix,superscriptaddress]{revtex4-1}
\usepackage{bm}
\usepackage{amsmath}
\usepackage{epsfig}
\usepackage{color}
\usepackage{natbib}
\usepackage{graphicx}
\usepackage{hyperref}
\usepackage{ifthen}
\usepackage{xstring}
\usepackage{graphicx}

\newcommand{\curv}{{\cal R}}

\definecolor{darkgreen}{cmyk}{0.85,0.2,1.00,0.2} 
\definecolor{purple}{cmyk}{0.5,1.0,0,0}

\begin{document}
\title{Generalized Slow Roll  for Non-Canonical Kinetic Terms}

\author{Wayne Hu}
\affiliation{Kavli Institute for Cosmological Physics, Department of Astronomy \& Astrophysics, University of Chicago, Chicago IL 60637}

\date{\today}

\begin{abstract}
\baselineskip 11pt
We show that the generalized slow-roll approach for calculating the power spectrum where the inflationary slow roll parameters are neither small nor slowly varying can be readily extended to models with non-canonical kinetic terms in the inflaton action.   
For example, rapid sound speed variations can arise in DBI models with features in the
warp factor leading to features in the power spectrum.   Nonetheless there remains a single source function for deviations that is simply related to the power spectrum.   Empirical constraints
on this source function can be readily interpreted in the context of features in the
inflaton potential or sound speed.   
\end{abstract}

\maketitle

%%%%%%%%%%%%%%%%%%%%%%%%%%%%%%%%%%%%%%%%%
\section{Introduction} \label{sec:intro}
%%%%%%%%%%%%%%%%%%%%%%%%%%%%%%%%%%%%%%%%%

Inflation with non-canonical kinetic terms in the scalar field action can arise in braneworld
inflation (e.g.~\cite{Silverstein:2003hf}).   Here the sound speed for fluctuations in the inflaton field can deviate from unity.  
The power spectrum and higher order $N$ point 
functions are usually computed under the slow roll or slow variation approximation
developed in \cite{ArmendarizPicon:1999rj,Garriga:1999vw,Chen:2006nt} where the sound
speed is assumed to be adiabatically varying.
 
As in the case of inflation with canonical kinetic terms, slow variation in the parameters
is not a requirement, neither for sufficient inflation nor for acceptable
power spectra with current constraints.    Rapid variation that is either localized in time or that has little cumulative
effect on the expansion history is allowed and can lead to interesting observational
signatures in the power spectrum and non-Gaussianity.  For example, these
can arise from features in the potential (e.g.~\cite{Starobinsky:1992ts,Chen:2006xjb}) or the warp factor of Dirac-Born-Infeld (DBI) inflation
\cite{Bean:2008na}.

While such cases can always be solved on a model by model basis, the generalized slow roll (GSR) 
approach  \cite{Stewart:2001cd} provides an efficient computational tool that also clarifies the relationship between
 cosmological observables and the model for the inverse problem.   In this {\it Brief Report},
we show that the technique can be readily extended to the case of non-canonical kinetic
terms.  With this extension, all of the results of the GSR formalism apply to non-canonical terms as well
\cite{Choe:2004zg,Dodelson:2001sh,Kadota:2005hv,Dvorkin:2009ne,DvoHu10a}.   In particular, there remains a single source function of the background evolution that describes even relatively large deviations from
the slow variation assumption in the power spectrum.

We briefly review how the background expansion history relates to the general inflaton action and the usual slow variation parameters in 
\S \ref{sec:background}.  In \S \ref{sec:GSR}, we
describe the GSR formalism for calculating the power spectrum when the slow variation parameters are neither small nor slowly varying.  We discuss applications  of the GSR formalism in \S \ref{sec:discussion}.

%%%%%%%%%%%%%%%%%%%%%%%%%%%%%%%%%%%%%%%%%
\section{Background Evolution}
\label{sec:background}
The
action 
\begin{equation}
S_{\phi} = \int d^4 x \sqrt{-g}\, p(X,\phi) ,
\end{equation}
for the inflaton $\phi$ 
provides a generalization of the canonical $p = X-V(\phi)$ case to arbitrary functions
of the kinetic term \cite{ArmendarizPicon:1999rj}
\begin{equation}
X =- {1 \over 2} \nabla^\mu \phi \nabla_\mu\phi .
\end{equation}
The scalar field behaves as a perfect fluid with pressure $p$ and
energy density
\begin{equation}
\rho = 2 X p_{,X} - p .
\end{equation}
The equations of motion for the background value of the field are given by the fluid
equation of motion
\begin{equation}
{d \ln \rho \over d\ln a} = -3 (1+p/\rho),
\end{equation}
or equivalently by the field equations 
\cite{Gordon:2004ez}
\begin{eqnarray}
{d \phi \over d\ln a} &=& {\chi \over H} , \nonumber\\
{d \chi \over d\ln a} &=& - 3 c_s^2 \chi - {1 \over H} {\rho_{,\phi} \over  \rho_{,X}},
\label{eqn:eom}
\end{eqnarray}
with $X = \chi^2 /2$.  Here the  sound speed is
\begin{equation}
c_s^2 = {p_{,X}  \over \rho_{,X}} = {\rho + p \over 2 X \rho_{,X}} = {p_{,X} \over p_{,X} + 2 X p_{,XX}} ,
\end{equation}
and the Hubble parameter obeys the usual Friedmann equations where we now assume that
the scalar field dominates the energy density during inflation
\begin{eqnarray}
H^2 & = & { \rho \over 3} , \nonumber\\
{d\ln H \over d\ln a} &=& -{3\over 2} (1 + p/\rho) \equiv -\epsilon_H .
\end{eqnarray}
Throughout we take units where $M_{\rm pl} = 1/\sqrt{8\pi G}=1$.

Inflation occurs when the slow roll parameter $\epsilon_H < 1$.   Sufficient inflation requires $\epsilon_H$ to remain small for many e-folds leading to conditions on the second slow roll parameter
\begin{eqnarray}
\eta_H = -\delta_1 &\equiv& \epsilon_H - { 1\over 2} {d\ln \epsilon_H \over d\ln a}.
\label{eqn:epsilon}
\end{eqnarray}
In the ordinary slow roll case, $\eta_H$ is also slowly varying and hence
sufficient inflation requires $\epsilon_H \ll 1$ and $|\eta_H | \ll 1$  around the e-folds when  large scale structure left the horizon. 

The GSR approach relaxes this condition by allowing the
third slow roll parameter 
\begin{eqnarray}
 \delta_2  &\equiv& \epsilon_H \eta_H + \eta_H^2 - {d\eta_H \over d\ln a} 
\end{eqnarray}
to become large and cause evolution in $\eta_H$.

By employing the equations of motion (\ref{eqn:eom}), these slow roll parameters
can be related to the evolution of the field $\phi(\ln a)$ through the functional
form of $p(X,\phi)$.  For a canonical kinetic term the result is 
\begin{eqnarray}
\left( { V_{,\phi} \over V} \right)^2  = 2\epsilon_H { (1-\eta_H/3)^2 \over (1-\epsilon_H/3)^2}, \nonumber\\
\left( { V_{,\phi\phi} \over V} \right) = {\epsilon_H + \eta_H - \delta_2/3 \over 1-\epsilon_H/3},
\end{eqnarray}
and returns the familiar relationship between $\epsilon_H$, $\eta_H$  and
the potential in the ordinary slow roll limit.

For non-canonical kinetic terms the general procedure of obtaining the evolution of the slow roll parameters from the equations of motion still holds but involves
other partial derivatives of $p(X,\phi)$ beyond $p_{,\phi}$ and $p_{,\phi\phi}$.   In particular, the sound speed enters
and so it is convenient 
 to introduce an additional slow variation parameter
\cite{Chen:2006nt}
\begin{equation}
\sigma_1 \equiv  {d \ln c_s \over d\ln a} .
\end{equation}
The ordinary slow variation  approximation would require $\epsilon_H$, $|\eta_H|$ and $|\sigma_1| \ll 1$
as well as constant to leading order.   In the GSR approach we also allow
$\sigma_1$ to evolve as monitored by a final parameter
\begin{equation}
\sigma_2 \equiv {d \sigma_1 \over d\ln a}=  {d^2 \ln c_s\over d\ln a^2 },
\end{equation}
bringing the number of slow variation parameters derived from the background solution to 
five.  

As a concrete example, consider DBI inflation
\cite{Silverstein:2003hf} where
\begin{equation}
p(X,\phi) = -{1\over F(\phi)} \sqrt{ 1- 2 F(\phi) X} - V(\phi)
\end{equation}
and hence
\begin{eqnarray} 
\epsilon_H(X,\phi) &=&  {3  X F(\phi) \over 1 + c_s(X,\phi) V(\phi)F(\phi)  } ,\nonumber\\
c_s^2(X,\phi) &=& 1 - 2 F(\phi) X, \nonumber\\
\rho =  3H^2 &=&  {1 \over F(\phi) c_s(X,\phi) } + V(\phi)\,.
\end{eqnarray}
Here $F(\phi)$ is related to the warp factor in brane inflation.

Taking the derivative of $\epsilon_H$ and employing the field equation ({\ref{eqn:eom}), we
obtain
\begin{eqnarray}
{d \ln \epsilon_H \over d\ln a} & =& {c_s \over \chi F^2 H (1+ c_s F V)} \Big\{ 
2 c_s F_{,\phi} + 2 F F_{,\phi} V 
\nonumber\\
&& - 2 F^2 ( 3  c_s \chi H + \chi^2 V F_{,\phi} + V_{,\phi} ) \nonumber\\
&&+ \chi^2 F^4 V ( 3 \chi H + c_s V_{,\phi})  \nonumber\\
&&+ F^3[-6 \chi H V + (\chi^2 - 2 c_s V) V_{,\phi}] \Big\},
\label{eqn:epsHevol}
\end{eqnarray}
which can be taken to define $\eta_H(\ln a)$.  Likewise the derivative of Eq.~(\ref{eqn:epsHevol}) can be used to define $\delta_2(\ln a)$ which in turn involves the
derivative of the sound speed or 
\begin{eqnarray}
\sigma_1 &=& {\chi \over F H} \Big[ - F_{,\phi} +  F^2 ( 3 \chi H + c_s V_{,\phi}) \Big] .
\end{eqnarray}
Note that the sound speed can change suddenly if there is a feature in either $F(\phi)$ or
$V(\phi)$ and $\sigma_2$ contains second derivatives of these functions.
For example $F(\phi)$ might contain steps \cite{Bean:2008na} from duality cascade 
\cite{Strassler:2005qs}.

\section{Generalized Slow Roll}
\label{sec:GSR}

Fluctuations in the inflaton field $\delta \phi$ in spatially flat slicing are related to the curvature fluctuations on
comoving (or constant field) slicing $\curv$ as 
$u = z \curv= -z(d\phi/d\ln a)^{-1} \delta\phi$, where
\begin{equation}
z \equiv { a (\rho+p)^{1/2} \over c_s H} .
\label{eqn:z}
\end{equation}
For any background evolution, regardless of values or evolution of the
five slow variation parameters, the field fluctuations obey
\cite{Garriga:1999vw}
\begin{equation}
\ddot u + c_s^2 k^2 u - {\ddot z \over z} u = 0 ,
\label{eqn:Mukhanov}
\end{equation}
where overdots are derivatives with respect to the conformal time $\eta$ to the
end of inflation, which we define as positive
and decreasing.  

Given that field oscillations will freeze out at sound horizon crossing rather
than horizon crossing,
it is useful to transform variables to
\begin{equation}
s \equiv
 \int c_s d\eta =  \int_{a}^{a_{\rm end}} { d a\over a} {c_s \over aH }
\end{equation}
and rescale the field variable
$y=\sqrt{2 k c_s } u$ to remove adiabatic effects from evolution in the sound speed.
Eq.~(\ref{eqn:Mukhanov}) then becomes
\begin{equation}
{d^2 y \over ds^2} + \left( k^2 - {2 \over s^2} \right) y = {g(\ln s) \over s^2} y ,
\label{eqn:yeqn}
\end{equation}
where 
\begin{equation}
g \equiv {f'' - 3 f' \over f} ,
\end{equation}
with $'  \equiv d/d\ln s$ and
\begin{equation}
f= 2\pi z c_s^{1/2} s = \sqrt{ {8\pi ^2 } {\epsilon_H c_s \over  H^2}} {a H s \over c_s}.
\label{eqn:f}
\end{equation}
Thus the field equation contains up to second derivatives in both $\epsilon_H$ and $c_s$
and hence involves the slow variation parameters up to $\delta_2$ and $\sigma_2$.

In these variables, the equation of motion for the field fluctuations are identical
for canonical and non-canonical kinetic terms.
For $ks \gg 1$ or deep inside the sound horizon, the $g$ term is suppressed by
  $(ks)^{-2}$ and the solutions are just
free plane  waves $\exp(\pm i ks)$.   The $\exp(+i ks)$ solution is the positive energy, Bunch-Davies 
vacuum, initial condition and is valid for any non-divergent evolution of the slow variation parameters.

For $ks \ll 1$ the curvature fluctuations freeze out
and the curvature power spectrum is given by
\begin{equation}
\Delta_\curv^2 \equiv {k^3 P_\curv \over 2\pi^2} = \lim_{ks \rightarrow 0} \left| { ks y \over f }\right|^2 .
\end{equation}
Note that in the lowest order slow variation approximation $f$ remains constant with
$ a H s/ c_s \rightarrow 1$ and 
$ks y \rightarrow 1$ so that  \cite{Garriga:1999vw}
\begin{equation}
\Delta_\curv^2 \approx f^{-2} \approx  {H^2 \over 8\pi^2 \epsilon_H c_s} .
\end{equation} 

Non-canonical kinetic terms require three simple generalizations of the GSR approach. The first is that features in field space are mapped onto features in $k$ space through the sound horizon $s$ rather than the particle horizon $\eta$.  
The second is that the source function $g$ of deviations from slow variation contains
derivatives of the sound speed.  Finally, the derivatives of 
$f$ are taken with respect to the sound horizon rather than conformal time.
 With these generalizations in mind, all of the
results proven for GSR with canonical kinetic terms apply to non canonical
kinetic terms as well \cite{Stewart:2001cd,Choe:2004zg,Dvorkin:2009ne}.

Briefly,
the GSR approach to solving the field equation (\ref{eqn:yeqn})
 is to consider the RHS as an external source with an iterative correction to the
 field value $y$.    To lowest order, we replace $y \rightarrow y_0$ where 
\begin{equation}
y_0 = \left( 1 + {i \over ks} \right) e^{i ks} ,
\end{equation}
is the solution with $g \rightarrow 0$ corresponding to the initial conditions above.  We then solve for the field fluctuation $y$ through the Green function technique.   Thus the GSR approximation has extended validity in that it allows
features in 
 $p(X,\phi)$ to cause relatively large changes in $X$ and the field dynamics through $c_s$ as long as the field position remains slowly varying.

To first order, the curvature power spectrum is given by
\begin{equation}
\ln \Delta_\curv^{2 \, (1)} = G(\ln s_{\rm min}) + \int_{s_{\rm min}}^\infty {d s\over s} W(ks) G'(\ln s) .
\end{equation}
where 
\begin{equation}
G = - 2 \ln f    + {2 \over 3} (\ln f )'   .
\end{equation}
and thus
\begin{equation}
G' = -2 (\ln f )' + {2 \over 3}  (\ln f )''  = {2 \over 3}  g - {2 \over 3} [(\ln f)']^2  ,
\label{eqn:Gprime}
\end{equation}
with the window function
\begin{equation}
W(u) = {3 \sin(2 u) \over 2 u^3} - {3 \cos (2 u) \over u^2} - {3 \sin(2 u)\over 2 u} .
\end{equation}
The addition of the term quadratic in $(\ln f)'$ to $g$  in Eq.~(\ref{eqn:Gprime}) 
guarantees that the power spectrum
is independent of the arbitrary epoch $s_{\rm min}$ after sound horizon crossing,
ensuring that the curvature remains constant thereafter \cite{Dvorkin:2009ne}.  
Note that the local slope \cite{Kadota:2005hv}
\begin{equation}
n_s -1 \equiv {d \ln \Delta_\curv^2 \over d\ln k}  \approx \int {d s \over s}  W'(ks) G'(\ln s).
\end{equation}
Since $\int d\ln u W' = -1$, $n_s - 1 = -G'$ for slowly varying $G'$.

To second order,
\begin{equation}
 \Delta_\curv^{2\, (2)} = \Delta_\curv^{2\, (1)}  \left\{ [ 1+ {1\over 4}I_1^2(k) + {1\over 2}I_2(k)]^2 + {1 \over 2}I_1^2(k) \right\}
 \label{eqn:second}
\end{equation}
where
\begin{eqnarray}
I_1(k) &=& { 1\over \sqrt{2} } \int_0^\infty {d s \over s} G'(\ln s) X(ks) , \nonumber\\
I_2(k) &=& -4 \int_0^\infty { d u \over u } [ X + {1\over 3} X' ] {f' \over f} F_2(u) ,
\end{eqnarray}
with 
\begin{equation}
F_2(u) = \int_u^\infty {d \tilde u \over \tilde u^2} {f' \over f},
\end{equation}
and
\begin{equation}
X(u) = {3 \over u^3} (\sin u - u \cos u)^2 .
\end{equation}
For cases where $f''/f$ controls the large deviations in $G'$, the dominant second order term is
$I_1$ and hence the power spectrum depends only on a single source function $G'$ through
two simple quadratures \cite{Dvorkin:2009ne}.

It is useful to relate the source function $G'$ to the 
five slow roll parameters.     Taking derivatives of $\ln f$, 
we obtain
\begin{eqnarray}
G'& =& {2 \over 3} (2 \epsilon_H -2\eta_H -\sigma_1) + 
{2 \over 3} ({ a H s \over c_s }-1)^2 \\
&& +  {2 \over 3} ({ a H s \over c_s} -1) (4 + 2\epsilon_H - 2\eta_H - \sigma_1) \nonumber\\
&& +{1 \over 3}  \left( { a H s \over c_s}\right)^2 
\Big[ 2 \delta_2 + 2\epsilon_H^2 - 2\eta_H - 2\eta_H^2 
\nonumber\\
&& - 3\sigma_1 + 2 \eta_H \sigma_1 
+ \sigma_1^2 - \epsilon_H ( 4\eta_H+\sigma_1) - \sigma_2\Big]. \nonumber
\end{eqnarray}
In the ordinary slow roll approximation one assumes $\epsilon_H \ll 1$, $|\eta_H| \ll 1$,
$|\sigma_1 |\ll 1$ and negligible $\delta_2$, $\sigma_2$.  Thus 
\begin{equation}
{a H s \over c_s } \approx 1+ \sigma_1 + \epsilon_H ,
\end{equation}
and so
\begin{equation}
G' \approx 4\epsilon_H - 2\eta_H + \sigma_1 = 1-n_s,
\end{equation} 
which returns the usual slow roll approximation for tilt generalized to non-canonical kinetic terms \cite{Chen:2006nt}.  Note that this derivation  makes it clear that the slow
variation of the sound speed during the many e-folds before sound horizon crossing does not have any observable impact on the power spectrum.

Another interesting limit is where we still require $\epsilon_H \ll 1$ and $|\sigma_1 |\ll 1 $ but allow $\eta_H$, $\delta_2$ and $\sigma_2$ to become large.   In this case
\begin{equation}
G' \approx  - 2\eta_H  - {2\over 3}\eta_H^2 + {2\over 3}\delta_2 - {\sigma_2 \over 3}.
\end{equation}
In particular $\delta_2$ and $\sigma_2$ can become large for a small number
of e-folds if there is a large second derivative term in the potential of a canonical
field or in the sound speed of a non-canonical field.   For example, in DBI inflation, steps in the
warp factor $F(\phi)$ produce very similar phenomenology to steps in $V(\phi)$ in the
canonical case as noted by \cite{Bean:2008na}.  

%--------------
\section{Discussion} \label{sec:discussion}
%--------------

We have shown that the generalized slow roll (GSR) approximation can be extended straightforwardly to
the case of non-canonical kinetic terms for the inflaton.   The extension involves only the remapping of epochs during inflation to wavenumber $k$ through the sound horizon rather
than the particle horizon and a generalization of the source function $g$ (or $G'$) of
deviations from slow variation to account for evolution in the sound speed.

The GSR formalism can be used to streamline the calculation of brane inflation models where the sound speed varies significantly across the observable e-folds.   An example is a
DBI model with steps in the warp factor \cite{Bean:2008na}.  We intend to examine these
and other models in a future work.  Our treatment can also be used to derive second order
corrections in the case where the 
slow variation parameters are both small and slowly varying \cite{Burrage:2011hd}.

The single source function $G'$ contains nearly all of the information from the power spectrum on single field inflation
with canonical or non-canonical terms even for strong evolution in the slow variation parameters \cite{Dvorkin:2009ne}.
Empirical constraints on $G'$ such as the percent level limits from the CMB acoustic
peaks  \cite{DvoHu10a} and weaker limits on large scale features \cite{Dvorkin:2011ui} can be directly reinterpreted in  
context of  braneworld models as constraints on variations in the sound speed 
rather than features in the inflaton potential.   Extensions of the GSR approach
to the bispectrum \cite{Adshead:2011bw} may also be useful for non-canonical models
where the non-Gaussianity can become large.

 \medskip
 \noindent {\it Acknowledgments}:   I thank Peter Adshead, Cora Dvorkin and
 Mark Wyman for useful conversations and Vinicius Miranda for pointing out a typo in the draft.  This work was supported by
 the KICP through grants NSF PHY-0114422 and NSF PHY-0551142, U.S.~Dept.\ of Energy contract
 DE-FG02-90ER-40560 and the David and Lucile Packard Foundation.

\vfill

%%%%%%%%%%%%%%%%%%%%%%%%%%%%%%%%%%%%%%%%%
\bibliography{Hu11}
%%%%%%%%%%%%%%%%%%%%%%%%%%%%%%%%%%%%%%%%%

\end{document}